Manuscript presented at the 75th International Astronautical Congress (IAC), Milan, Italy, 14-18 October 2024.

# Tailoring Infra-Red Filters for Global Mapping of Enceladus' Thermal Environment

**Duncan G. Lyster[†*], Dr. Carly J. A. Howett[*], Prof. Neil E. Bowles[*], Dr. Keith Nowicki[*], Rory Evans[*], Dr. Tristram J. Warren[*]**

[*] *Atmospheric, Oceanic and Planetary Physics (AOPP), University of Oxford, United Kingdom*
[†] *Corresponding Author* duncan.lyster@physics.ox.ac.uk

**Abstract**

We present the results of an initial study into the feasibility and design of the Enceladus Thermal Mapper (ETM) instrument, designed to provide detailed temperature data for Enceladus, an icy moon of Saturn identified by the Cassini Mission as a prime candidate for astrobiological study due to its subsurface ocean and active cryovolcanism. Building on heritage from previous missions including TechDemoSat-1, Mars Climate Sounder, and Lunar Trailblazer, ETM aims to measure temperature variations across Enceladus' surface, including at the poles during the extreme cold of winter, when they can reach as low as 45 K. Where Cassini provided limited high-resolution flyby data of Enceladus due to its Saturn-centric orbit, ETM, in orbit around the icy moon will offer continuous high-resolution thermal mapping with a 384 x 288-pixel sensor, capable of 80 m/pixel resolution from 150 km altitude. This will improve constraints on the Enceladus' global conductive heat loss, and better characterise and contextualise its active South Polar Terrain. To optimise the instrument's filters for thermal emission measurements, a digital model of the filter assembly was developed. This lightweight, low-power instrument is capable of significantly advancing the understanding of Enceladus' geological behaviour and evaluating its potential habitability by providing comprehensive temperature data essential for interpreting its evolution.

**Keywords:** Enceladus, thermal mapping, cryovolcanism, infrared filters, astrobiology, thermal properties

**Acronyms/Abbreviations**

Technology Readiness Level (TRL)
Lunar Thermal Mapper (LTM)
Enceladus Thermal Mapper (ETM)
Composite Infrared Spectrometer (CIRS)
Instantaneous Field-of-View (iFOV)
Infrared (IR)
South Polar Terrain (SPT)
Signal to Noise Ratio (SNR)
Time Delay Integration (TDI)

## 1. Introduction

Enceladus, at only 500 km across, is one of Saturn's smallest moons, but it stands out as one of the most interesting bodies in the solar system from an astrobiological and geophysical perspective. The Cassini mission provided our first close look at Enceladus, identifying it as one of the few bodies known to harbour liquid water. The warm 'Tiger Stripe' fractures at the South Pole and the cryovolcanic plumes jetting into space are particularly significant for understanding subsurface ocean conditions.

NASA's Decadal Survey identified this moon as a high-priority target for future missions, with one key recommendation being the Enceladus Orbilander [1]. This mission concept combines both orbiter and lander capabilities to conduct close-up studies of Enceladus' surface and subsurface. The New Frontiers 5 opportunity also lists Enceladus as a possible target, such a mission is likely to include Enceladus orbits – including multiple low passes, allowing new insights that were impossible with Cassini due to its orbital constraints around Saturn. One possibility, used as the main guide for this report, is an orbit at 150 km altitude.

The University of Oxford's multispectral thermal instrument is a high Technology Readiness Level (TRL) instrument, combining heritage from previous incarnations as the Lunar Thermal Mapper (LTM) [2] on NASA's Lunar Trailblazer mission (launching 2024), MIRMIS [3] - the hyper spectral and multispectral remote sensing instrument for ESA's Comet Interceptor mission (due to launch in 2029), as well as TechDemoSat-1 and Mars Climate Sounder.

As a low mass (~4.5 kg), low power (~6 W average, ~12 W peak) instrument, Oxford's multispectral thermal mapper is an ideally suited auxiliary payload for a large mission to Enceladus after some modifications. At 9.5 AU, Enceladus' temperatures range from winter nights of 30 K, to Tiger Stripe fracture temperatures of ~180 K.

The purpose of this investigation was to produce initial recommendations for the filters to be used in a modified filter assembly aimed at maximising the scientific output of this instrument, the Enceladus Thermal Mapper (ETM). The prolonged duration of such a mission, as well as the intensity of Saturn's radiative environment present additional challenges which are not discussed here, including requiring fully redundant, shielded electronics and power.

### *1.1 Enceladus' thermal environment*

Enceladus presents a new challenge for Oxford's multispectral mapping instrument. Previous missions at ~1 AU had sufficiently radiative targets to use narrow spectral bandpass channels with low total power transmission for surface composition mapping. At 9.5 AU, however, the Saturn system receives only ~1% of the sunlight that reaches Earth, this combined with Enceladus' high albedo means that the winter night-time temperatures can get as low as 30 K. The warmest surface temperatures are believed to reach ~ 200 K [4] and are found along Enceladus' South Pole's 'Tiger Stripe' fractures, believed to be linked to endogenic tidal heating.

Temperature variations due to changes in illumination differences across day/night, summer/winter, and even eclipse observations are a vital source of information when attempting to characterise the material properties of the surfaces of planetary bodies. Transient heating by the Tiger Stripe's cryovolcanic plumes on Enceladus may also be responsible for heating the area around the fractures. In each of these cases, the surface temperature, and the rate at which it cools down gives us information about the properties of the regolith (e.g., albedo, density, thermal conductivity etc.) [5].

### *1.2 Scientific goals*

The thermal emission profile of Enceladus can tell us a huge amount about the surface regolith, composition, and subsurface thermal emission. Cassini performed global thermal mapping in the mid-IR (12.5 to 16 μm) at a spatial resolution of 25 km [6]. From these observations it has proved difficult to disentangle endogenic thermal emission from re-radiated solar heat. Mapping at higher spatial resolution covering a greater range of local times will allow for tighter constraints on global thermal emission, and may indicate whether the emission (and thus possibly the sub-surface ocean) can be sustained in the long-term - a key criterion for astrobiological potential. Measurements of the south pole during the night-time in winter are of particular importance, as this will minimise the heat signature due to solar heating, allowing the nearly isolated contribution from endogenic heat production to be measured.

The rate of diurnal heating and cooling constrains the thermal inertia of the top few centimetres of the surface; combining this with spectral information about composition can reveal how consolidated the ice is, giving indications of its density and thermal conductivity. Over longer timescales, seasonal temperature variations reach ~m levels below the surface, and brief eclipse occultations as the moon passes through the shadow of Saturn affect only the temperature of the top few millimetres, which can tell us about the crystal structure of the ice [7].

Multi-spectral thermal mapping instruments have previously been used in tandem with spectrometers to map mineral composition at the surfaces of planetary bodies [2]. The extreme cold of Enceladus makes this challenging, but this report also investigates the feasibility of such operations for mapping salinity on Enceladus and the relevant trade-offs for this instrument.

### *1.3 Digital instrument model for filter assembly calculations*

We will discuss the newly developed model of the instrument filter assembly, which creates a framework for systematically assessing and optimising infrared bandpass filter assemblies to meet scientific goals. Strategically chosen filter profiles will enable the determination of the full seasonal and diurnal range of Enceladus' thermal emission, allowing for precise temperature measurements with a goal of improving constraints on global thermal emission due to tidal heating. The suitability of different filter assemblies for different science goals will be discussed.

## 2. The Enceladus Thermal Mapper Instrument

Between its Composite Infrared Spectrometer (CIRS), and Visual Infrared Mapping Spectrometer (VIMS) Cassini achieved high spectral and spatial resolution, with its highest spatial resolution detectors (CIRS focal planes 3 and 4) having 10 pixels, each with an iFOV of 0.273 mrad [1]. At its lowest 74-km-altitude pass this allowed for measurements with a pixel 'footprint' of just 38 x 214 m [4]. However, due to the limited flyby nature of Cassini much of Enceladus was left without this high spatial resolution thermal mapping. The University of Oxford's multi-band radiometric instrument operates 384 cross-track line scanning pixels, each with an iFOV of 0.540 mrad (see Tab. 1 for further details). The instrument has space for 15 wavelength bands and operates as a 384 x 288 pixel push-broom sensor. Preliminary mission concepts anticipate flying this instrument in orbit around Enceladus at an altitude of 150 km, allowing ETM to globally map Enceladus at 80 m/pixel resolution, with a track 31 km wide.

### *2.1 Microbolometer array sensors*

Each pixel in the sensor is a microbolometer, an uncooled thermal radiation measurement device that works on the principle of radiative equilibrium. The instrument's optics focus IR radiation from the target onto a low mass, low thermal inertia absorber. This absorber heats up or cools down to the colour temperature of the target within the range selected by the bandpass filters. The absorber is thermally isolated from the readout electronics, and connected only by an electrically conductive bridge (see Fig. 1). The temperature dependence of the absorber's conductivity allows the temperature to be measured, giving a readout. Readings from many pixels can be combined in a push broom fashion to build up a global view.

| | |
|---|---|
| **iFOV** | 0.540 mrad per pixel |
| **Mass** | 4.5 kg (with margin), 3.8 kg (actual pFM mass) |
| **Power** | 6 W (average), ~12 W (peak) |
| **Volume** | 262.5 x 214 x 100 mm (excluding MLI blankets) |
| **Internal storage** | 1 GB |
| **Power connector** | RS422 |
| **Calibration** | Internal calibration target and space view for radiometric calibration |
| **Nominal voltage** | 28V |

Table 1: Instrument overview

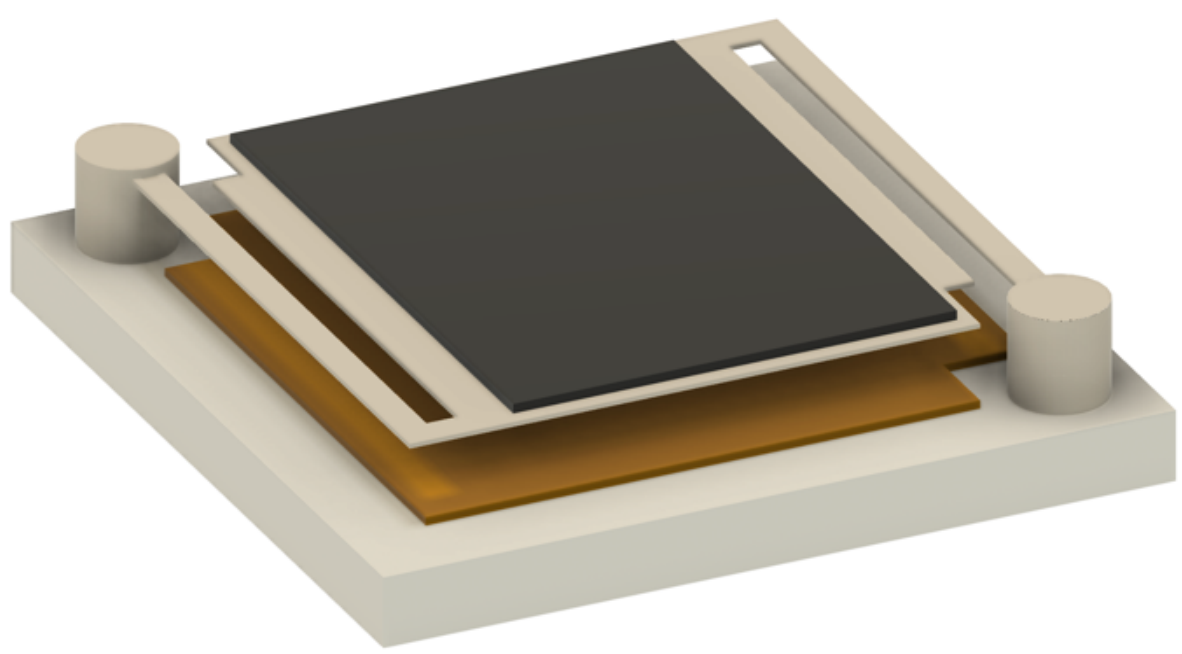

Fig. 1. Diagram of a pixel component from a microbolometer array showing the absorbing surface, electrically conducting bridge connecting the absorber to the readout electronics, and the mirrored surface below the absorber.

## *2.2 Infrared filters*

The IR filters discussed in this report fall into three categories: low pass filters, metal mesh filters, and multilayer interference filters.

Low pass filters truncate the incoming radiation at a cut-off frequency, with the frequency depending primarily on the properties of the material used for the filter. For example, polyethylene doped with ionic salts is a popular choice, having in-band transmittance as high as 0.95, and good high-frequency rejection – see e.g. [11]. Due to their cut-off nature, low pass filters are best suited for low intensity, long wavelength radiation, and in this study are recommended for improving the signal-to-noise ratios in measurements of the very coldest areas of the target body.

Metal mesh filters use stacks of a capacitive and inductive mesh grids to select the target bandpass. They are well-suited to applications in the 20-200 µm wavelength range, and are proposed for filter assemblies optimised for temperature measurements in the 45-180 K range.

Multilayer interference-based filters comprise many thin alternating layers of materials with different refractive indices designed to reflect and transmit chosen wavelengths. The Oxford Infrared Multilayer Laboratory specialises in producing bandpass filters in the 2-45 µm range. These are particularly well suited to narrow bandpass filters, such as those used on LTM for surface composition mapping. In this study, model multilayer bandpass filters at 6.6-9.2 µm and 8.5-9.2 µm are evaluated for their effectiveness in mapping salinity, while broad bandpass CdTe and ZnS filters are proposed for filter assemblies optimised for investigating the relatively warm South Polar Terrain (SPT) due to their alignment with the 14.5 µm Wien peak for a black body at 200 K.

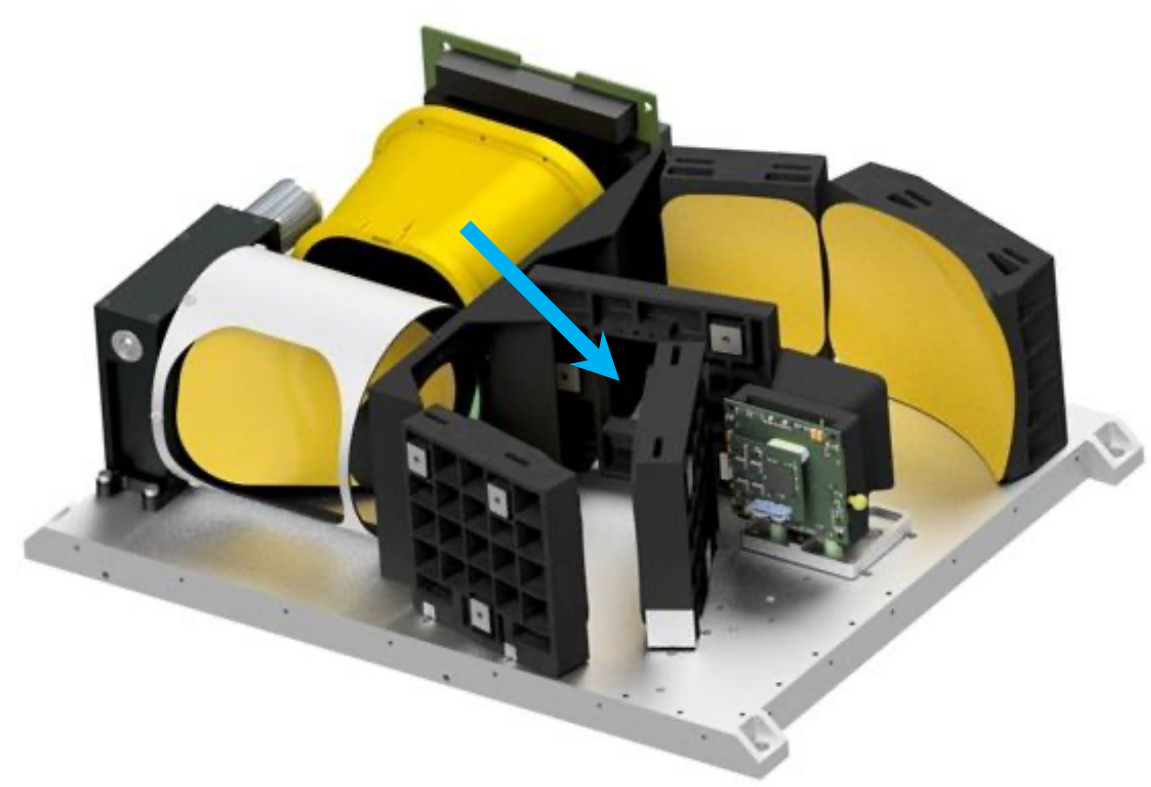

Fig. 2. A digital render of the Enceladus Thermal Mapper instrument. The arrow shows the location of the filter assembly.

## *2.3 Filter assembly geometry*

The filter assembly is mounted at the focal plane of the instrument's optics, as shown in Fig. 2. The filters are individually produced, then attached together before being fixed over the sensor as seen in Fig. 3. The mounting mechanism requires a 3-pixel gap to be left between multilayer interference filters, and a 5-pixel gap between mesh filters, or a mesh filter and an interference filter. The filters can be aligned along the long or the short edge of the sensor, prioritising either the maximum number of TDI pixels possible, or the maximum number of cross track pixels. For the ETM, we opt for the shorter cross-track case to maximally benefit from extended along track time delay integration.

### *2.4 Selection criteria for bandpass filters*

**Temperature Measurement:** As the primary goal of the ETM is to produce global temperature mapping coverage, it must be capable of accurately capturing thermal emissions across the range of temperatures expected on Enceladus. The low radiance at the lowest temperatures means broad bandpass filters with high in-band transmittance are prioritised to maximise SNR.

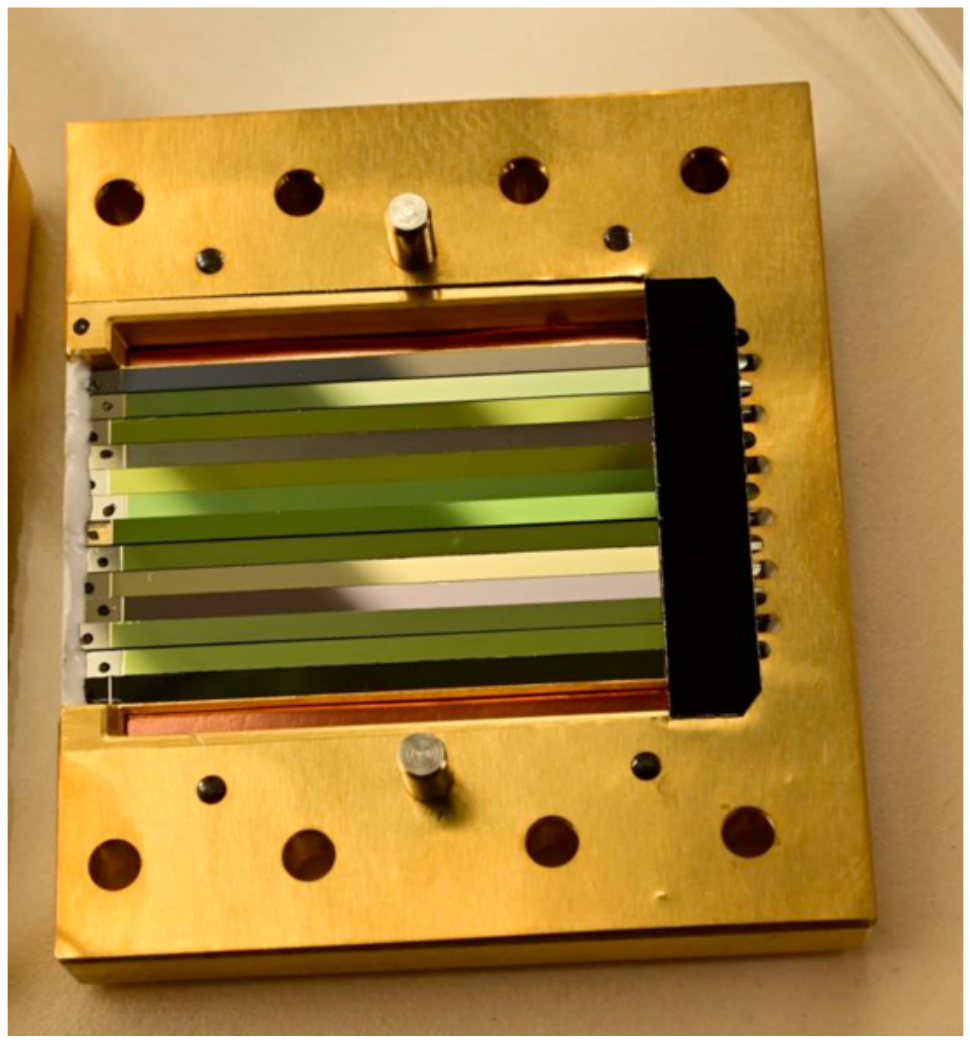

Fig. 3. The filter assembly from the LTM. Note that the ETM will likely have just 4 to 6 vertically aligned filters, rather than the horizontally aligned 15 filters shown here.

**Identification of Spectral Characteristics:** Multispectral mapping instruments have a rich history of being used to map mineral composition of planetary bodies by using signal ratios from filters carefully selected to coincide with spectral features of minerals of interest (see e.g. [2] [3]). To determine the feasibility of such a study on Enceladus we investigated a design aimed at identifying NaCl on the surface of the icy moon. NaCl was specifically targeted because it has been observed in the plumes ejected from Enceladus' subsurface ocean [10]. By necessity the chosen filters must be narrow enough to focus on a specific spectral feature, reducing the range of collected wavelengths and thus the transmitted power.

**Physical constraints:** With 384 along track pixels to allocate, bandpasses need to be traded off against each other to optimise overall performance. Introducing more filters also increases the number of pixels that are unused due to being covered by joins between filters.

### *2.5 Mathematical basis*

The quality of a detector is often characterised by the specific detectivity, $D*$, defined as:

$$D^* = \frac{\sqrt{A_d}}{NEP} \tag{1}$$

This parameter combines the effects of detector area $A_d$ and Noise Equivalent Power, $NEP$ (in W Hz$^{1/2}$) allowing for a comparison of different detectors on an equal footing. To calculate the transmitted power, and SNR per pixel for each bandpass we integrate the blackbody spectrum over the selected filter profile. SNR can be calculated from equation (1):

$$SNR = D^* \sqrt{t} \sqrt{A_d} \Omega_d \int \tau \epsilon_S B(\tilde{v}, T_s) d\tilde{v} \tag{2}$$

Where $D*$ is $10^9$ cm Hz$^{1/2}$ W$^{-1}$, the measured value for the detector used in LTM. $\sqrt{A_d}$ is 35 µm, the side length of a pixel on the microbolometer array. $\Omega_d$ is 0.663 sr, the solid angle subtended by the pixel. $\tau$ is the total throughput of the system accounting for each reflection, and the absorption of the detector. $\epsilon_S$ is the emissivity of the source (assumed to be 1 for Enceladus). $t$ is the integration time, $t = t_{pixel} N_{pixel}$ where $N_{pixel}$ is the number of TDI pixels in the along-track width of the filter, and $t_{pixel}$ is the integration time for one pixel, calculated to be 0.6 s, based on an iFOV of 41 m/pixel from an orbital altitude of 150 km in the scenario investigated. $B(\tilde{v}, T_s)$is the spectral radiance of the source at temperature $T_S$ and calculated from the Planck formula.

### *3.2.1 Composition mapping*

To map the surface composition, we can use the ratio between two bandpass filters, a 'continuum' band, and a 'feature' band (see Fig. 4). To calculate the SNR required for a detection we can take the ratio of the transmitted power for the two filters for an iFOV containing none of the chemical (in this case a black body), and that for an iFOV containing just the chemical (NaCl in this case). Setting the detection level to 3σ, we calculate the SNR required for these filters at 200 K. We can derive the integration time required for this SNR, and from this we are able to retrieve the number of passes over a certain point required for a detection of NaCl. Note that this method gives the Limit of Detection (LOD) threshold required for a detection under 'perfect' conditions, i.e., a surface of pure NaCl.

### *2.6 Python code outline*

The analysis was conducted using an open-source Python script, which implements the calculations for thermal filter analysis. The script is designed to be modular, allowing for easy modification of instrument parameters and filter configurations. This flexibility enables quick iteration and optimization of filter designs for different mission scenarios. The full script is available online at `github.com/duncanLyster/thermal-mapper-filters`.

*2.7 Preliminary filter assembly profiles*

Initial calculations were performed with individual filters, over a range of integration times and target temperatures to create four candidate filter assembly profiles, each with a different intended specialism as a proof of concept. See Fig. 4. These could be adjusted as required to meet a mission's science requirements.

**Low temps:** A filter set aimed at best possible SNR at the lowest temperatures expected on Enceladus. This assembly makes use of a low pass filter, to collect the longest wavelengths with the least transmission loss, as well as three mesh filters covering a range up to 25 µm. This maximises the SNR for the lowest temperatures but reduces the ability to characterise the blackbody curve, meaning for example that the ability to identify small hotspots within a generally cold FoV is reduced.

**High temps:** Making use of two interference filters, this assembly extends visibility into the Wien peak of the highest temperatures encountered by Cassini at the focus of the tiger stripe fractures. This comes at the cost of losing the ability to characterise the coldest areas of the icy moon.

**Salinity:** Based on a 'best case' for the detection of salt, this assembly devotes half the available TDI pixels to just two spectral channels, one centred on a dip in the spectrum of NaCl, and one on the continuum. This assembly was used to investigate whether it may be possible to map salinity variations across the surface of Enceladus by using ETM in combination with a high-resolution spectrometer, and by comparing the ratio of the signals received by each channel.

**Generalist:** Based on initial investigations with each of the above a filter assembly was designed to attempt to find a compromise between the high and low temperatures optimised assemblies. Composition channels were not used.

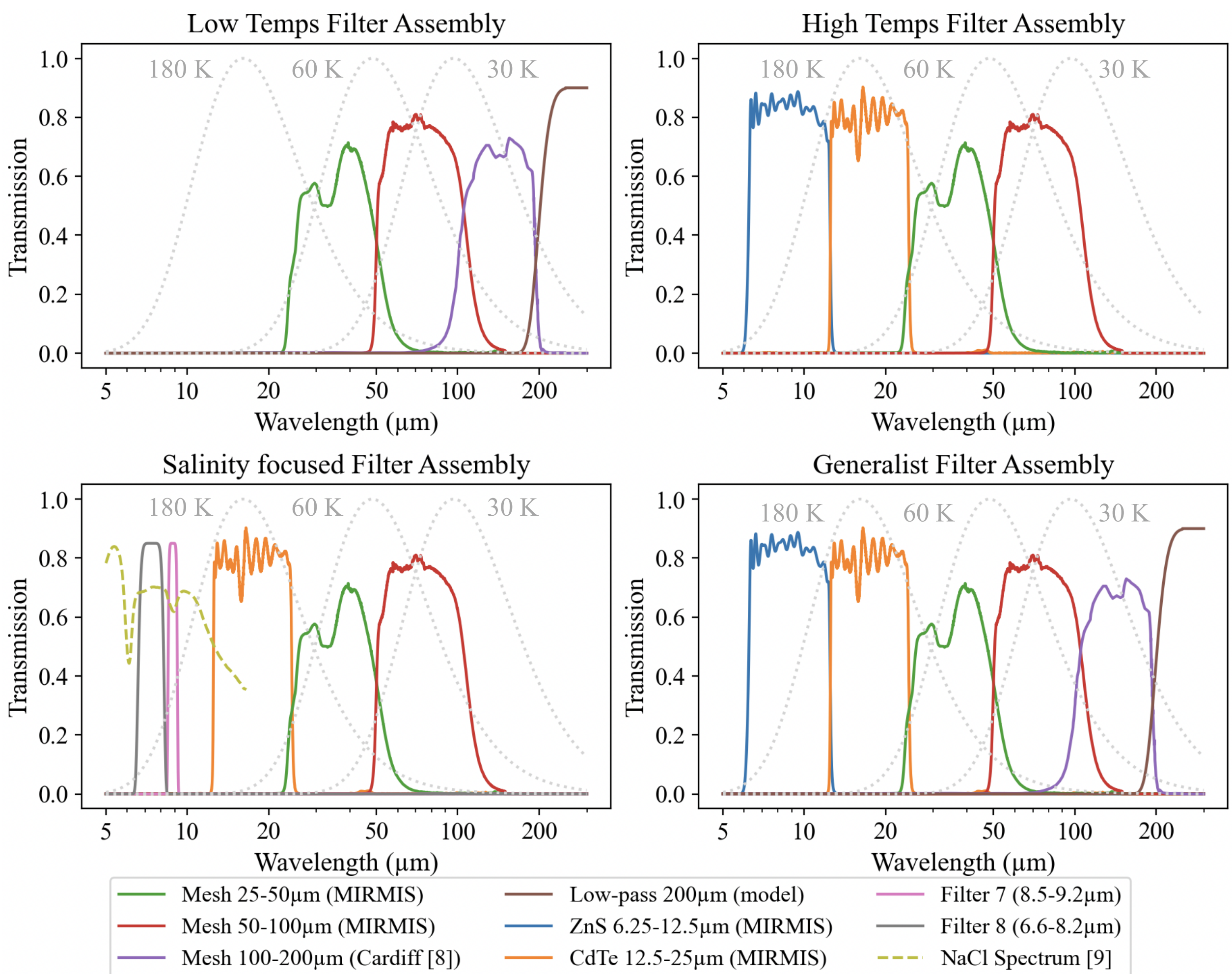

Fig. 4. The filter profile assemblies investigated. Normalised blackbody curves are included to show alignment of filters with the relevant emission profile.

## 3. Results & Discussion

Table 1 presents the calculated SNRs for each filter for the given width (number of TDI pixels) for one pass. It is possible to increase the SNR by coadding observations from multiple orbits over the same features, so these figures should be taken as a baseline for observations.

| Filter | Description | Width (px) | 30K | 60K | 180K |
|---|---|---|---|---|---|
| **Low Temps Filter Assembly** | | | | | |
| 3 | Mesh 25-50µm | 92 | 0.26 | 55.1 | 5030 |
| 4 | Mesh 50-100µm | 92 | 4.63 | 115 | 1730 |
| 5 | Mesh 100-200µm | 92 | 5.73 | 42.1 | 299 |
| 6 | Low-pass 200µm | 92 | 2.61 | 9.89 | 43.6 |
| **High Temps Filter Assembly** | | | | | |
| 1 | ZnS 6.25-12.5µm | 92 | 0.00 | 0.00 | 2180 |
| 2 | CdTe 12.5-25µm | 92 | 0.02 | 3.13 | 8680 |
| 3 | Mesh 25-50µm | 92 | 0.26 | 55.1 | 5030 |
| 4 | Mesh 50-100µm | 92 | 4.63 | 115 | 1730 |
| **Salinity Focused Filter Assembly** | | | | | |
| 7 | 8.5-9.2µm (dip) | 60 | 0.00 | 0.00 | 161 |
| 8 | 6.6-8.2µm (flat) | 120 | 0.00 | 0.00 | 224 |
| 2 | CdTe 12.5-25µm | 60 | 0.02 | 2.53 | 7010 |
| 3 | Mesh 25-50µm | 60 | 0.21 | 44.5 | 4060 |
| 4 | Mesh 50-100µm | 60 | 3.74 | 92.7 | 1390 |
| **Generalist Filter Assembly** | | | | | |
| 1 | ZnS 6.25-12.5µm | 60 | 0.00 | 0.00 | 1760 |
| 2 | CdTe 12.5-25µm | 60 | 0.02 | 2.53 | 7010 |
| 3 | Mesh 25-50µm | 60 | 0.21 | 44.5 | 4060 |
| 4 | Mesh 50-100µm | 60 | 3.74 | 92.7 | 1390 |
| 5 | Mesh 100-200µm | 60 | 4.62 | 34.0 | 242 |
| 6 | Low-pass 200µm | 60 | 2.10 | 7.98 | 35.2 |

Table 2. Calculated signal to noise ratios for a set of recommended filter assembly options. SNRs are colour coded as follows: red <1, orange 1-10, light green 10-50 and dark green >50. The width given is the number of TDI pixels allocated to each filter within the assembly.

### *3.1 Composition mapping*

Analysis of the 'Salinity' assembly (filters 7 & 8) reveals a potential opportunity for using multispectral imaging for compositional mapping even at these low temperatures. For a 3σ detection of NaCl with the selected salinity bandpass filters assuming a pure NaCl surface, an SNR of 52 is required, this is achievable at ~160 K. By allocating 50% of the sensor's pixels to filters matched to this spectral feature, our instrument achieves an SNR of 161 after a single pass at 180 K (Fig. 5). While temperatures this high are only found in the immediate region surrounding the tiger stripe fractures, it is possible to co-add observations from multiple passes to expand the range of usefully mapped composition into colder regions.

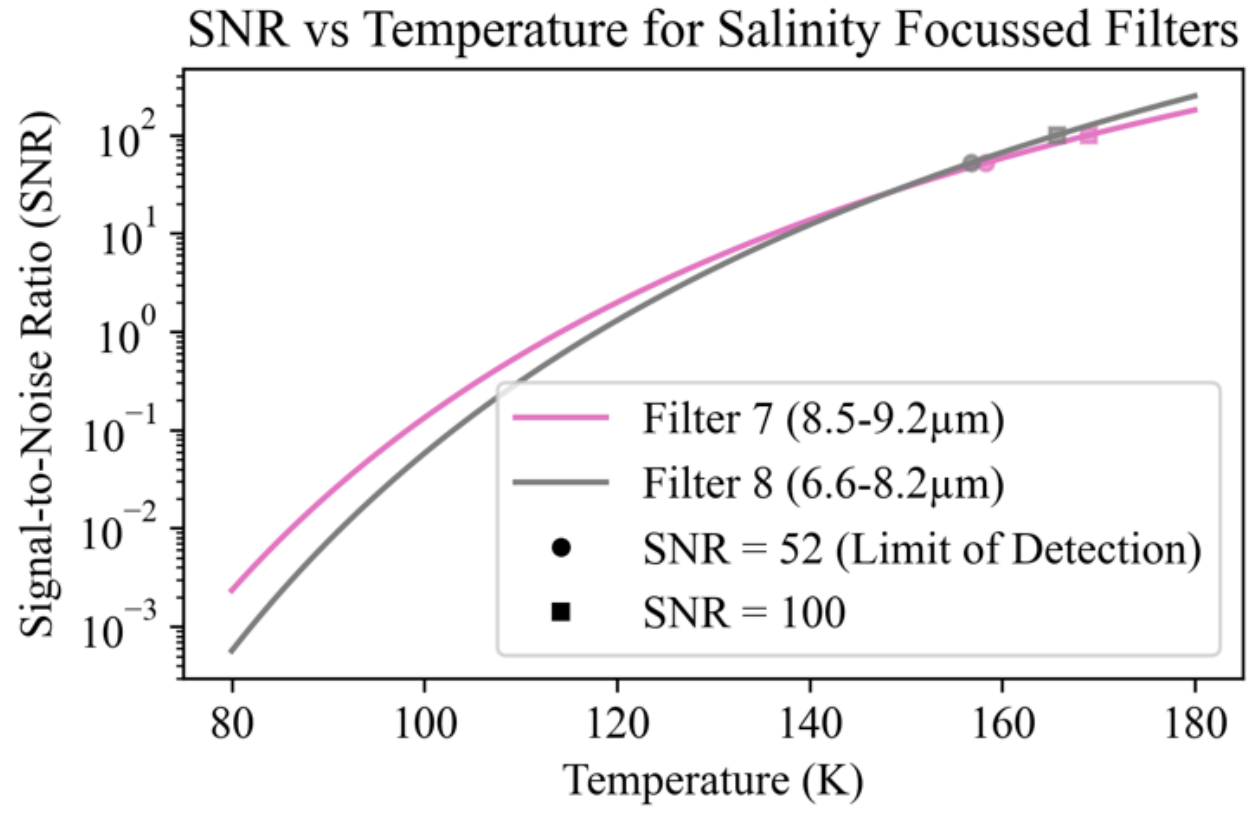

Fig. 5. Signal to Noise ratios for the filters used to investigate composition mapping. The SNRs were calculated for filter widths of 60 pixels, and 120 pixels for filters 7 and 8, respectively.

### *3.2 Filter assembly recommendations*

While the low and high temps filter assemblies showed improved fidelity within their chosen temperature ranges, the loss in SNR to either was small when they were combined into the generalist filter assembly. The most significant reduction in that case is for the 200 µm low pass filter with a 30 K target. The low temps filter assembly gives a SNR of 2.6 (assuming a 92 TDI pixel array), which decreases to 2.1 for the generalist filter assembly (60 pixels). Coadding signals from multiple passes will effectively increase the SNR, potentially mitigating the reduction seen in the generalist filter assembly and making it a viable option depending on the mission's specific scientific objectives.

### *3.3 Limitations and further research*

One issue that falls outside the scope of this investigation is a comprehensive diffraction analysis. The longest target wavelengths are greater than the pixel size, meaning thermal mapping resolution will be limited by diffraction rather than the iFOV of the instrument for the coldest regions. Understanding this in detail requires investigation with an optical model of the full instrument and is outside the scope of this work.

Some other important physical considerations have not been addressed due to the theoretical nature of this work. Chief amongst these is heating of the sensor and instrument optics by spacecraft components, and internal instrument electronics, which may set a cut-off on the lowest observable temperatures at Enceladus. A structural thermal model (STM) would offer some insights, but deploying this instrument in the extreme thermal environment of Saturn would require a comprehensive calibration and testing campaign.

The spectral properties of the ice on the surface of Enceladus are not well understood, calibrating the instrument with different phases of water ice at representative temperatures would be highly valuable.

The potential for compositional mapping demonstrated, raises the possibility of including more spectral filters, with the goal of mapping a variety of minerals on the surface. This will come at the cost of SNR of the thermal channels, but depending on the mission configuration and STM this may be a favourable trade-off.

## 4. Conclusions

The ETM's design, with its high-heritage, high-TRL, low mass, low power consumption, and high-spatial resolution capabilities, aligns well with future mission concepts to Enceladus. This instrument sits ready to play a key role in addressing outstanding questions about the moon's subsurface ocean, cryovolcanic activity, and astrobiological potential.

This investigation identified specific filter profile combinations tailored for different temperature ranges and scientific goals, highlighting a generalist filter assembly as the most balanced option for comprehensive thermal mapping. On-going work will constrain the instrument's diffraction limits and spacecraft-induced thermal interference through further laboratory testing and calibration.

This initial study for the Enceladus Thermal Mapper (ETM) instrument demonstrates the potential this small instrument holds for advancing our understanding of Enceladus' thermal environment and geological behaviour. Building on heritage from previous missions, Oxford's thermal instrument requires only minor adaptations to provide continuous high-resolution thermal mapping of this Saturnian moon. ETM will improve constraints on tidal heating effects and aid in evaluating the moon's potential habitability.

**Acknowledgements**

This work was supported by a PhD studentship from the Science and Technology Facilities Council (STFC), and ETM development is supported by UK Space Agency through its Bilateral Program.